# A near-field radiative heat transfer device

John DeSutter[1], Lei Tang[2], and Mathieu Francoeur[1,*]

[1]Radiative Energy Transfer Lab, Department of Mechanical Engineering, University of Utah, Salt Lake City, UT 84112, USA

[2]Department of Mechanical Engineering, University of California, Berkeley, Berkeley, CA 94720, USA

**Recently, many works have experimentally demonstrated near-field radiative heat transfer (NFRHT) exceeding the far-field blackbody limit between planar surfaces[1–15]. Due to the difficulties associated with maintaining the nanosize gaps required for measuring a near-field enhancement, these demonstrations have been limited to experiments that cannot be implemented into actual applications. This poses a significant bottleneck to the advancement of NFRHT research. Here, we describe devices bridging laboratory-scale measurements and potential NFRHT engineering applications in energy conversion[16,17] and thermal management[18–20]. We report a maximum NFRHT enhancement of ~ 28.5 over the blackbody limit with devices made of millimeter-sized doped silicon (Si) surfaces separated by vacuum gap spacings down to ~ 110 nm. The devices capitalize on micropillars, separating the high-temperature emitter and low-temperature receiver, manufactured within micrometer-deep pits. These micropillars, which are ~ 4.5 to 45 times longer than the nanosize vacuum spacing where radiation transfer takes place, minimize parasitic heat conduction without sacrificing device structural integrity. The robustness of our devices enables gap spacing visualization via scanning electron microscopy (SEM) prior to**

---
*email: mfrancoeur@mech.utah.edu

**performing NFRHT measurements. Direct gap spacing characterization is critical for transitioning NFRHT research from laboratory-scale experiments to applications.**

In the near field (i.e., subwavelength vacuum gap spacing), tunneling of evanescent modes allows for radiative heat transfer to exceed Planck's far-field blackbody limit by orders of magnitude[21]. While NFRHT research is primarily motivated by potential performance enhancement in energy conversion and thermal management technologies, NFRHT devices that can be implemented into engineering applications are yet to be realized. Precision alignment systems[1–6,17,20] are well-suited for laboratory demonstration of NFRHT, but integration of such systems into actual applications is not feasible. Measurements of NFRHT between surfaces separated by micro/nanosize vacuum gap spacings supported by particle[7,8] or microfabricated[9–11,18] spacers and via microelectromechanical systems[12–15,19] have been performed. However, significant thermal conduction[8–11,18] between the emitter and receiver greatly reduces the effectiveness of any potential devices capitalizing on NFRHT. Fragile and intricate structures are difficult to manufacture and characterize[15]. Devices requiring external forces[12–15,19] to maintain desired nanosize gap spacings further characterization difficulties and greatly complicate practical implementation. Finally, surfaces of microsize dimensions[12,13,19] severely limit the total radiative heat exchange. We circumvent these limitations by fabricating and characterizing bonded devices suitable for potential engineering applications of NFRHT. These devices independently support their own gap spacing (standalone), have surfaces with macroscale dimensions, minimize parasitic heat conduction, and their structural integrity enables gap spacing visualization via SEM.

A NFRHT device, manufactured using standard micro/nanofabrication techniques as detailed in Methods and Supplementary Fig. 1, is shown in Fig. 1a. It consists of a high-temperature emitter and low-temperature receiver made of Si, with boron doping of $\sim 4.6\times10^{19}$ cm$^{-3}$, separated by low thermal conductivity SU-8 3005 micropillars (0.2 Wm$^{-1}$K$^{-1}$)[22] having diameters of either $\sim$ 20 or 30 µm. Both emitter and receiver are 525-µm-thick, have surface area of 5.2 × 5.2 mm$^2$, and are characterized by RMS surface roughness of less than 0.2 nm as provided by the manufacturer (Silicon Valley Microelectronics). Approximately 4.5-µm-deep, 215-µm-diameter pits are etched into the emitter substrate where the micropillars are manufactured. The pits enable devices with micropillars significantly longer than the nominal gap spacing, $d$, between the emitter and receiver, thus minimizing the contribution of parasitic conduction to the total heat rate[23]. The micropillar and pit areas respectively cover 0.01% and less than 1.2% of the total surface of the device. A 100-µm-wide frame is etched into both the emitter and receiver substrates to prevent particle contamination at the edges of the device, due to dicing and handling, from interfering with the desired gap spacing. On the emitter side, the frame is etched to the same depth as the pits while the receiver frame is $\sim$ 8-µm-deep. The resulting separation distance between the emitter and receiver along the edges of the device is greater than 12.5 µm, which is much larger than most particles. After meticulously cleaning the emitter and receiver (see Methods), the micropillars are bonded to the receiver surface. The robustness of the fabricated NFRHT devices enables imaging of the gap spacing $d$ via SEM. Figure 1b shows SEM images of gap spacing at the four corners of a device with $d \approx 380$ nm. SEM images allow direct gap spacing characterization prior to performing heat transfer measurements.

Heat transfer measurements are conducted using the setup shown in Fig. 1c located inside a vacuum chamber ($P < 5\times10^{-4}$ Pa) housed in a class 1000 cleanroom tent. The emitter is heated by a thermoelectric heat pump (Custom Thermoelectric, 00701-9B30-22RU4) while the receiver temperature is held constant at $\sim$ 300 K via a thermoelectric cooler (TETechnology, VT-31-1.0-1.3). The heat rate across the device, due to radiation heat transfer ($Q_{rad}$) and conduction heat transfer through the micropillars ($Q_{cond}$), is measured with a custom built $10 \times 10$ mm$^2$ heat flux meter (HFM) from FluxTeq (PHFS-JD10). Two thermistors (Selco, LSMC700A010KD002), one embedded in a 0.5-mm-thick, $5 \times 5$ mm$^2$ copper heat spreader located between the heat pump and emitter ($T_h$) and one embedded in an identical heat spreader placed directly under the receiver ($T_l$), are used to measure the temperature difference across the device. Additional 0.5-mm-thick, $10 \times 10$ mm$^2$ copper heat spreaders surround the HFM to ensure uniform flux across the meter. Contact resistance is minimized by applying thermal grease (Arctic Silver Ceramique 2) at all interfaces. The resulting thermal resistance between $T_h$ and $T_e$, and $T_l$ and $T_r$ is ~ 4.75 KW$^{-1}$, where $T_e$ and $T_r$ are the emitter and receiver temperatures adjacent to the vacuum gap (see Supplementary Section 1). Heat transfer measurements are calibrated using 1.1-mm-thick, $5 \times 5$ mm$^2$ samples of soda lime glass with known thermal conductivity of 0.94 Wm$^{-1}$K$^{-1}$; HFM calibration is detailed in Supplementary Section 1 and Supplementary Fig. 2.

Fig. 2a compares theoretical and experimental near-field radiative heat flux, $q_{rad}$, for six devices with varying vacuum gap spacings and temperature differences $\Delta T = T_e - T_r$ ($T_r = 300 \pm 0.5$ K) ranging from ~ 5 to 100 K (radiation from the pits and frame is not included). SEM images of the gap spacing at the four corners of each device are provided in Supplementary Fig. 3. The gap spacing range provided in the legend of Fig. 2a for a specific device is determined from the SEM images, and a thermal and structural analysis of the device (see Supplementary Section 2). For example, the gap spacing at each corner of the device leading to the largest radiative flux is estimated to be 92, 109, 114, and 122 nm at room temperature. Theoretical radiative flux is calculated using fluctuational electrodynamics[21,24] (FE). The radiative flux associated with a specific device is computed via the Derjaguin approximation using the four gap spacings derived from SEM imaging and the measured bow of the Si substrates (see Methods and Supplementary Figs. 4-6). The colored theoretical bands arise from uncertainty in the vacuum gap spacing extracted from SEM images, Si doping levels, and micropillar height and diameter (see Methods for uncertainty analysis). Theoretical and experimental trends are in good agreement, and the radiative flux measured for all devices exceeds the far-field blackbody limit. A maximum conduction contribution of ~ 22 to 35% is estimated for the largest gap device (874-982 nm), while the conduction contribution reaches a minimum of ~ 1.9 to 4.1% for the smallest gap device (92-122 nm). The radiative heat transfer coefficient, $h_{rad}$, and the enhancement over the blackbody limit, $E_{BB}$, are shown in Fig. 2b as a function of the vacuum gap spacing for a temperature difference of 70 ± 2 K. The device with the smallest gap spacing is characterized by a $h_{rad}$ value of ~ 247 $Wm^{-2}K^{-1}$, which falls within the upper range of forced convection with gases. This leads to a substantial radiative transfer enhancement over the blackbody limit, $E_{BB}$, of approximately 28.5. Unprocessed heat rate data that includes radiative transfer between the emitter and receiver separated by a gap spacing $d$, radiative transfer from the bottom of the pits and frame (recessed areas) to the receiver, and conduction through the micropillars are provided in Supplementary Fig. 7. When partial or full contact between the emitter and receiver is forced, the measured heat rate greatly exceeds heat transfer for the smallest gap device (see Supplementary Fig. 8), which is an additional proof that heat transfer in the devices is mediated by NFRHT.

Near-field enhancement is explained by analyzing the radiative flux as a function of the angular frequency, $\omega$, and wavevector parallel to the emitter and receiver surfaces, $k_\rho$, for devices with the largest and smallest gap spacing, and a temperature difference of 70 K (see Fig. 3a). In transverse magnetic polarization, doped Si supports surface plasmon-polaritons (SPPs) characterized by large parallel wavevectors $k_\rho$ exceeding the material light line Re$(n)k_0$, where $n$ is the refractive index of doped Si and $k_0$ is the magnitude of the wavevector in vacuum[25]. For reference, the SPP dispersion relation ($\omega^+$ and $\omega^-$) in the Si-vacuum-Si configuration is plotted in Fig. 3a (see Methods). In the electrostatic limit, the largest contributing parallel wavevector to the flux for doped Si supporting SPPs in the infrared is estimated as $k_\rho \approx d^{-1}$.[26] For the smallest gap device (calculated here as $d = 110$ nm), the flux is dominated by SPPs evanescent in both vacuum and Si with $k_\rho$ greatly exceeding the material light line (~ 82% of the flux is due to SPPs). This is also observed in Fig. 3b, in which the monochromatic flux for the smallest gap device is maximum near the resonant frequency of a Si-vacuum interface $\omega_{SPP} \approx \omega_p / \sqrt{\varepsilon_\infty + 1} =$ $1.77 \times 10^{14}$ rad/s, where $\omega_p$ is the plasma frequency (= $6.29 \times 10^{14}$ rad/s) and $\varepsilon_\infty$ is the limiting value of the dielectric function at high frequency (= 11.7) (see Supplementary Section 3). SPP resonance of a Si-vacuum interface is derived by assuming the materials are lossless, which explains the small discrepancy with FE predictions in Fig. 3b. For the largest gap device (calculated here as $d = 1000$ nm), the flux is still dominated by evanescent modes in vacuum (~ 73% of the flux is due to evanescent modes). The largest contribution comes from frustrated modes (~ 53% of the flux), characterized by parallel wavevectors $k_0 < k_\rho <$ Re$(n)k_0$, that are propagating in Si but evanescent in vacuum (see Fig. 3a). This leads to a broadband enhancement of the flux, as opposed to the narrowband enhancement mediated by thermal excitation of SPPs obtained with the smallest gap device (see Fig. 3b).

The biggest challenge in transitioning NFRHT from laboratory-scale experiments to engineering applications is fabricating standalone, structurally robust devices while minimizing the relative contribution of conduction to the total heat rate. Our devices overcome this challenge by manufacturing micropillars, separating the emitter and receiver, inside micrometer-deep pits. Extending micropillar height to a few micrometers while keeping the gap spacing, $d$, in the range ~ 100 to 1000 nm substantially increases the thermal resistance by conduction, $R_{cond}$, between the emitter and receiver. For example, conduction heat transfer is reduced by a factor of ~ 42 when comparing a 110-nm-gap device without pits to the same device with 4.5-μm-deep pits. For our smallest gap device leading to a NFRHT enhancement of ~ 28.5 beyond the blackbody limit, the contribution of conduction to the total heat rate would increase from ~ 1.9% with pits to 45% without pits. Despite the large enhancement of NFRHT, a pit-free-device would be unusable for applications such as thermophotovoltaic energy conversion where heat conduction is detrimental to device performance[27]. Micropillars with relatively large diameters, (here, 20 to 30 μm) ensuring device structural integrity without having the drawback of large parasitic heat conduction, can be fabricated by capitalizing on micrometer-deep pits. Such structural integrity enables direct gap spacing characterization via SEM, which is critical in assessing the quality of our NFRHT devices. To our knowledge, this is the first time nanoscale gap spacings have been imaged in the context of NFRHT across macroscale surfaces with both lateral dimensions exceeding 1 mm.

Measurement of NFRHT across macroscale planar surfaces at a gap spacing as small as ~ 110 nm has never been reported. Here, the impact of surface area cannot be understated. While it is easier to maintain sub-100-nm vacuum gap spacing between microsize surfaces due to simpler parallelization and decreased likelihood of surface defects and contamination, the radiative heat rate is severely limited. For instance, for a temperature difference of ~ 10 K, the radiative heat rate in our smallest gap device is ~ 300 times larger than the heat rate obtained across microsize planar surfaces separated by a gap spacing of ~ 25 nm[4]. Note that maintaining a vacuum gap spacing on the order of 50 nm or less with our devices would be very challenging due to substrate bow.

In vacuum, a blackbody provides an upper limit for radiative heat transfer in the wavevector range $k_\rho < k_0$. Therefore, the only way to transfer radiation exceeding the blackbody limit across a vacuum gap is by tunneling evanescent modes with $k_\rho > k_0$. This is, indeed, possible in the near field.[21] In the far field, evanescent modes cannot contribute to radiative transfer and wavevectors are limited to $k_\rho < k_0$[28-30]. The NFRHT devices proposed here are, therefore, critical for the development and implementation of applications capitalizing on radiation transfer exceeding the blackbody limit.

In summary, we successfully fabricated and characterized NFRHT devices with gap spacings from ~ 1000 nm down to ~ 110 nm separating millimeter-sized surfaces of doped Si. Our singular design capitalizes on long micropillars, manufactured inside micrometer-deep pits, minimizing parasitic heat conduction without sacrificing structural integrity. These devices constitute a critical step towards realizing potential NFRHT applications in energy conversion and thermal management. The NFRHT devices described here cannot be operated at temperatures higher than ~ 450 K[31] due to instability of SU-8. However, by keeping the same design and by adjusting the fabrication process (e.g., hybrid SU-8/$SiO_2$ micropillars), we anticipate that the proposed devices can sustain temperature differences exceeding 1000 K.

**Methods**

**Device fabrication and cleanliness**. The main steps required for fabricating NFRHT devices are shown in Supplementary Fig. 1. The 100-mm-diameter Si wafers used for both the emitter and receiver, purchased from Silicon Valley Microelectronics, are characterized by bow smaller than 4.5 µm. Throughout the entire process, emitter and receiver surfaces are only exposed to class 100 or 1000 cleanroom environments. The emitter is fabricated by first spinning on a thick (~ 15 µm) layer of AZ 9260 photoresist (PR) (step 1). Pit and frame pattern generation is achieved by exposing the AZ 9260 PR to UV light shadowed by a photomask in a Suss MA1006 mask aligner (step 2). The pits and frame are then etched to a depth of approximately 4.5 µm using a $CF_4O_2$ reactive ion etch (RIE) in an Oxford Plasmalab 80 for 1 hour and 50 minutes (step 3). The frame is implemented into device design to minimize the impact of debris from dicing and handling. The dicing saw can produce significant particle contamination that is primarily concentrated near the sample edges. Therefore, the recessed frames prevents the majority of this debris from interfering with the desired gap spacing, *d*. The masking AZ 9260 layer is then removed using acetone, isopropanol (IPA), and a short $O_2$ RIE. SU-8 3005 permanent photoresist is then spin-coated with two different spin settings (step 4). Spin 1 is for 8 seconds at 500 rpm with a ramp rate of 100 rpm/s. Spin 2 is for 35 seconds at 2650 rpm with a ramp rate of 300 rpm/s. This is immediately followed by a soft bake at 95˚C for 135 seconds. 5.5 to 6.5-µm-thick SU-8 micropillars are patterned (step 5) via exposure to 120 mJ of UV radiation shadowed by a photomask. To produce micropillars with flat surfaces, a post exposure bake (PEB) at 70˚C for 1 minute preludes a PEB at 95˚C for 1 minute. The SU-8 is developed for 2 minutes. The emitter pattern is then cut into 5.2 × 5.2 mm$^2$ die using a Disco DAD641 dicing saw. To help avoid Si debris while dicing, a thick (10 − 15 µm) protective AZ 9260 PR layer is deposited onto the emitter wafer (step 6) and dicing tape is then adhered on the protective layer. The receiver frame is patterned and diced using a similar procedure to that of the emitter (see steps 7 to 10 in Supplementary Fig. 1).

After dicing, the tape and AZ 9260 are removed in a sonicated acetone bath for 1 minute (step 11). Once the tape is removed, the samples are immediately moved to a second sonicated acetone bath for 5 minutes. This is followed by 1 minute IPA and deionized (DI) water sonicated baths. Micropillar height is characterized using a Tencor P-20H profilometer. To achieve desired height and uniformity, the micropillars are selectively etched in an $O_2$ plasma while using a suspended shadowmask (step 12). Shorter micropillars are shadowed resulting in a slower etch rate. The iterative process of profilometer characterization and $O_2$ plasma etching is carried out until micropillars have the desired height and a uniformity less than 20 nm.

The surfaces of both the emitter and receiver must be pristine prior to bonding. If there is debris on the surfaces when viewed through an Olympus MX51 microscope once the micropillars have the desired height and uniformity, the edges of the top surfaces of the emitter and receiver are wiped with a cleanroom cloth (CONTEC Polywipe-C) soaked in IPA. The samples are then sprayed with acetone, IPA, and DI water to remove any additional debris the cloth may have left. This is another iterative process involving sample inspection in the Olympus microscope and the wipe/spray cleaning procedure that is undertaken until no visible particles are detected on the emitter and receiver surfaces. This is a delicate process as wiping the micropillars must be avoided. The emitter and receiver are then aligned using a square alignment fixture and bonded in an oven for 30 minutes at 200°C (step 13). No additional pressure is applied to the device during the bonding process.

**Heat transfer calculations.** NFRHT is modeled using FE[21,24]. The propagating, $q_{rad}^{prop}$, and evanescent, $q_{rad}^{evan}$, components of the radiative flux are calculated as follows for two infinite planes separated by a vacuum gap spacing $d$:

$$q_{rad}^{prop} = \frac{1}{4\pi^2} \int\limits_0^\infty d\omega [\Theta(\omega,T_e) - \Theta(\omega,T_r)] \int\limits_0^{k_0} dk_\rho k_\rho \sum_{\tau=\text{TE,TM}} \frac{\left(1-\left|r_{0e}^\tau\right|^2\right)\left(1-\left|r_{0r}^\tau\right|^2\right)}{\left|1-r_{0e}^\tau r_{0r}^\tau e^{2i\,\text{Re}(k_{z0})d}\right|^2} \tag{1}$$

$$q_{rad}^{evan} = \frac{1}{\pi^2} \int\limits_0^\infty d\omega [\Theta(\omega,T_e) - \Theta(\omega,T_r)] \int\limits_{k_0}^\infty dk_\rho k_\rho e^{-2\,\text{Im}(k_{z0})d} \sum_{\tau=\text{TE,TM}} \frac{\text{Im}(r_{0e}^\tau)\,\text{Im}(r_{0r}^\tau)}{\left|1-r_{0e}^\tau r_{0r}^\tau e^{-2\,\text{Im}(k_{z0})d}\right|^2} \tag{2}$$

where the subscripts 0, $e$, and $r$ respectively refer to vacuum, emitter, and receiver. In Eqs. (1) and (2), $\Theta(\omega,T)$ is the mean energy of an electromagnetic state, $k_{z0}$ is the component of the vacuum wavevector perpendicular to an interface, and $r_{0e,r}^\tau$ is the Fresnel reflection coefficient at the vacuum/emitter ($e$) or vacuum/receiver ($r$) interface in polarization state $\tau$.

The evanescent component of the radiative flux includes frustrated modes and SPPs. Separate evanescent contribution to the radiative flux is obtained by performing the integration over the parallel wavevector in the range $k_0 < k_\rho < \text{Re}(n)k_0$ for frustrated modes, and for $k_\rho > \text{Re}(n)k_0$ for SPPs[32].

The radiative flux used for generating the results in Fig. 2a and Supplementary Fig. 7 is the sum of Eqs. (1) and (2). The data in Fig. 3a is generated by solving Eqs. (1) and (2) per unit angular frequency, $\omega$, and per unit parallel wavevector, $k_\rho$ (i.e., both integrations are dropped). The spectral radiative flux in Fig. 3b is produced by solving Eqs. (1) and (2) per unit angular frequency, $\omega$ (i.e., the integration over $\omega$ is dropped).

For a specific device, the radiative flux, $q_{rad}$, is calculated using the Derjaguin approximation[33], where the emitter and receiver are discretized into sub-surfaces characterized by uniform gap spacings. The Derjaguin approximation is applicable since the radius of curvature of the emitter and receiver, due to substrate bow, is much larger than the gap spacing. The measured bow of the doped Si substrates is smaller than ~ 25 nm (see Supplementary Figs. 4 and 5). For the 223-291 nm, 485-508 nm, 627-681 nm, and 874-982 nm devices, substrate bow has a negligible impact on the radiative flux. For these devices, stable and accurate radiative flux prediction is obtained by assuming that the gap spacing at each of the four corners derived from SEM imaging is uniform over a quarter of the emitter and receiver surfaces. For example, for the 223-291 nm device and a temperature difference of 50 K, the radiative flux vary by less than 1.3% when calculated by discretizing the emitter and receiver into sixteen sub-surfaces as opposed to four sub-surfaces. For the smallest gap device (92-122 nm), gap spacing variation due to substrate bow may have a non-negligible impact on the predicted radiative flux. As such, the radiative flux is calculated using gap spacings derived from SEM imaging and two-dimensional (2D) topographic mapping of doped Si substrate bow (see Supplementary Fig. 5). Stable and accurate radiative flux prediction for the smallest gap device is obtained by discretizing the emitter and receiver into nine sub-surfaces. Note that the bow of the 100-mm-diameter Si wafers decrease by ~ 4.5% when the temperature increases from ~ 300 K to ~ 400 K (see Supplementary Fig. 6). Therefore, variation of substrate bow as a function of temperature is neglected when calculating the radiative flux for the smallest gap device.

To calculate the radiative flux between the recessed areas (bottom of pits and frame) in the emitter and receiver, Eqs. (1) and (2) are used again, but with a gap spacing $t_{mp} = t_{pit} + d_{avg}$ for the pits and $t_{frame} = 12.5 \ \mu m + d_{avg}$ for the frame (see Fig. 1a), where $d_{avg}$ is the average gap spacing of the four measured corners. These equations assume a view factor of unity, which is an excellent approximation for the emitter-receiver portion of the device separated by a vacuum gap spacing, $d$. It is less accurate for the recessed areas where 2D effects may be relevant. The area of the pits and frame accounts for less than 9% of the total device surface area. For the largest gap device (874-982 nm), ~ 97% of the radiative heat rate is due to radiation exchange across the gap spacing, $d$. The largest gap device is more impacted by the pit and frame radiative transfer than any of the other devices investigated. Therefore, accounting for potential 2D effects is clearly not necessary.

One-dimensional, steady-state conduction through the SU-8 micropillars with thickness $t_{mp}$ is considered. This is justified by the fact that the micropillar temperature is uniform in the direction parallel to the Si surfaces. A temperature-independent thermal conductivity of 0.2 Wm$^{-1}$K$^{-1}$ for SU-8 is used in the calculations[22]. Contact resistances at the SU-8/Si interfaces are neglected since SU-8 reflows and fills voids during the bonding process.

**Uncertainty analysis**

*Experimental data*. Each experimental point consists of the average value of a set of data recorded by the HFM every second for at least two minutes once steady state is reached. The distribution uncertainty associated with a set of data is calculated by taking two standard deviations of the mean. The accuracy uncertainty of 5% is provided by the HFM manufacturer. These uncertainties are added together to obtain the total heat rate uncertainty. The heat rate is 0.447 ± 0.023 W for the case of largest uncertainty (smallest gap device, largest temperature difference) and 0.0028 ± 0.0004 W for the case of smallest uncertainty (largest gap device, smallest temperature difference).

Twenty-four measurements of thermal grease resistance ($R_{grease}$) between the hot-side thermistor, $T_h$, and $T_e$ and the cold-side thermistor, $T_l$, and $T_r$ (see Fig. 1c) were taken (see Supplementary Section 1). The uncertainty in these measurements is the difference between the maximum and minimum recorded values. Thermal grease resistance uncertainty has the largest influence on temperature uncertainty, which also considers the $\pm$ 0.1 K accuracy of each thermistor and the $\pm$ (0.2 % + 1 $\Omega$) accuracy of the LCR meter (BK Precision 889B). The temperature difference is 78.6 $\pm$ 1.8 K for the case of largest uncertainty (smallest gap device, largest temperature difference) and 9.2 $\pm$ 0.2 K for the case of smallest uncertainty (largest gap device, smallest temperature difference).

*Theoretical predictions*. The colored bands for theoretical predictions (see Supplementary Fig. 7) arise from uncertainty in the gap measured from the SEM images, uncertainty in the Si doping concentration determined from bulk resistivity measurements using a four-point-probe, and uncertainty in the amount of conduction varying with micropillar diameter and height. Since Fig. 2a only includes radiative flux, the theoretical bands in this case arise only from uncertainty in SEM images and Si doping concentration. Conduction uncertainty is accounted for in Fig. 2a in the uncertainty range of the experimental data since theoretical conduction is subtracted from experimental measurements.

The uncertainty in the SEM measurements is due to image resolution at the gap edges. The uncertainty is determined by measuring the maximum and minimum possible gap spacing between which the actual gap spacing exists.

Doping uncertainty arises from the discrepancy in measured values using a four-point-probe. For the entire batch of wafers, the largest and smallest measured doping concentrations are $4.9{\times}10^{19}$ cm$^{-3}$ and $4.3{\times}10^{19}$ cm$^{-3}$, respectively.

The uncertainty associated with conduction heat transfer includes ± ~ 0.5 μm in micropillar diameter derived from Keyence microscope images and the micropillar height uncertainty obtained from SEM images and profilometry measurements (Tencor P-20H) of pit depth ($t_{mp}$ = $t_{pit}$ + $d$).

The upper (lower) curve of each colored band consists of the smallest (largest) possible gap spacing at each corner based on SEM measurements, doping concentration producing the largest (smallest) heat flux, and largest (smallest) possible amount of conduction.

For example, the lowest part of the theoretical band at a temperature difference of 20 K for the smallest gap device in Supplementary Fig. 7 is calculated by first determining the largest possible gap spacing (derived from SEM images) at each of the four corners. The device radiative flux is computed via the Derjaguin approximation assuming a doping concentration of $4.9 \times 10^{19}$ cm$^{-3}$. Conduction is added using the minimum estimated micropillar diameter of 29.2 μm and the maximum estimated micropillar height of 4769 nm ($t_{mp}$ = $t_{pit,max}$ + $d_{avg,max}$ = 4658 nm + 119 nm = 4769 nm, where $t_{pit,max}$ and $d_{avg,max}$ are, respectively, the maximum possible pit depth and average of the largest possible gap spacing). Radiation from the recessed areas is finally included in the theoretical values. However, uncertainty in $t_{pit}$ and $d_{avg}$ is negligibly small such that radiation from the recessed areas has no impact on uncertainty.

**SPP dispersion relation**. The dielectric function of doped Si is described by the following Drude model[34]:

$$\varepsilon(\omega) = \varepsilon_\infty - \frac{\omega_p^2}{\omega(\omega + i\gamma)} \tag{3}$$

where $\varepsilon_\infty$ is the limiting value of the dielectric function at high frequency, $\omega_p$ is the plasma frequency and $\gamma$ is the scattering rate (see Supplementary Section 3 for dielectric function model). The SPP dispersion relation in the Si-vacuum-Si configuration is plotted in Fig. 3a by neglecting losses ($\gamma = 0$), and by assuming that the emitter and receiver have the same dielectric function calculated at a temperature of 370 K. Note that the dielectric function model of doped Si is temperature-dependent. This temperature-dependence is taken into account when calculating the radiative flux with Eqs. (1) and (2). Due to SPP coupling within the vacuum gap spacing, the dispersion relation splits into antisymmetric, $\omega^+$, and symmetric, $\omega^-$, modes that are respectively determined by numerically solving the following equations[35]:

$$\tanh\left(\frac{k_{z0}d}{2}\right) = -\frac{k_z}{k_{z0}\varepsilon} \tag{4}$$

$$\tanh\left(\frac{k_{z0}d}{2}\right) = -\frac{k_{z0}\varepsilon}{k_z} \tag{5}$$

where $k_z$ is the wavevector component perpendicular to the surface in the emitter/receiver. In the electrostatic limit (i.e., large parallel wavevector $k_\rho >> k_0$) where SPP coupling within the vacuum gap spacing is negligible, both the antisymmetric and symmetric modes converge to the resonant frequency of a Si-vacuum interface[35]:

$$\omega_{SPP} \approx \frac{\omega_p}{\sqrt{\varepsilon_\infty + 1}} \tag{6}$$

**Data availability**. The data that support the findings of this study are available from the corresponding authors upon reasonable request.

**Code availability**. The computer codes that support the findings of this study are available from the corresponding authors upon reasonable request.

## Acknowledgements

The authors acknowledge financial support from the National Science Foundation (Grant No. CBET-1253577). This work was performed in part at the Utah Nanofab sponsored by the College of Engineering, Office of the Vice President for Research, and the Utah Science Technology and Research (USTAR) initiative of the State of Utah. The authors appreciate the support of the staff and facilities that made this work possible. This work also made use of University of Utah shared facilities of the Micron Technology Foundation Inc. Microscopy Suite sponsored by the College of Engineering, Health Sciences Center, Office of Vice President for Research, and the Utah Science Technology and Research (USTAR) initiative of the State of Utah.

## Authors contributions

This work was conceived by J.D. and M.F. Design, fabrication, and testing of the device, as well as the associated numerical simulations, were performed by J.D. under the supervision of M.F. Calibration of the experimental system was done by J.D. and L.T. under the supervision of M.F. The manuscript was written by J.D. and M.F with comments provided by L.T.

## Competing financial interests

The authors declare no competing financial interests.

**a**

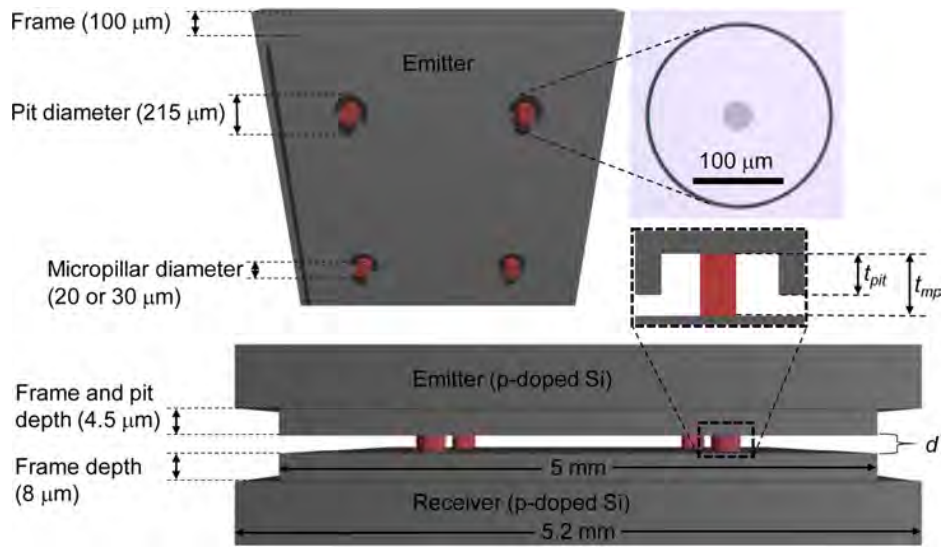

Frame (100 μm)

Emitter

Pit diameter (215 μm)

100 μm

Micropillar diameter
(20 or 30 μm)

$t_{pit}$

$t_{mp}$

Emitter (p-doped Si)

Frame and pit
depth (4.5 μm)

Frame depth
(8 μm)

5 mm

$d$

Receiver (p-doped Si)

5.2 mm

**b**

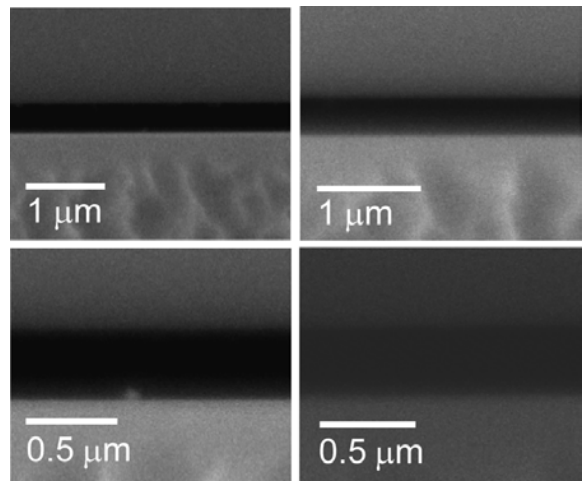

1 μm

1 μm

0.5 μm

0.5 μm

**c**

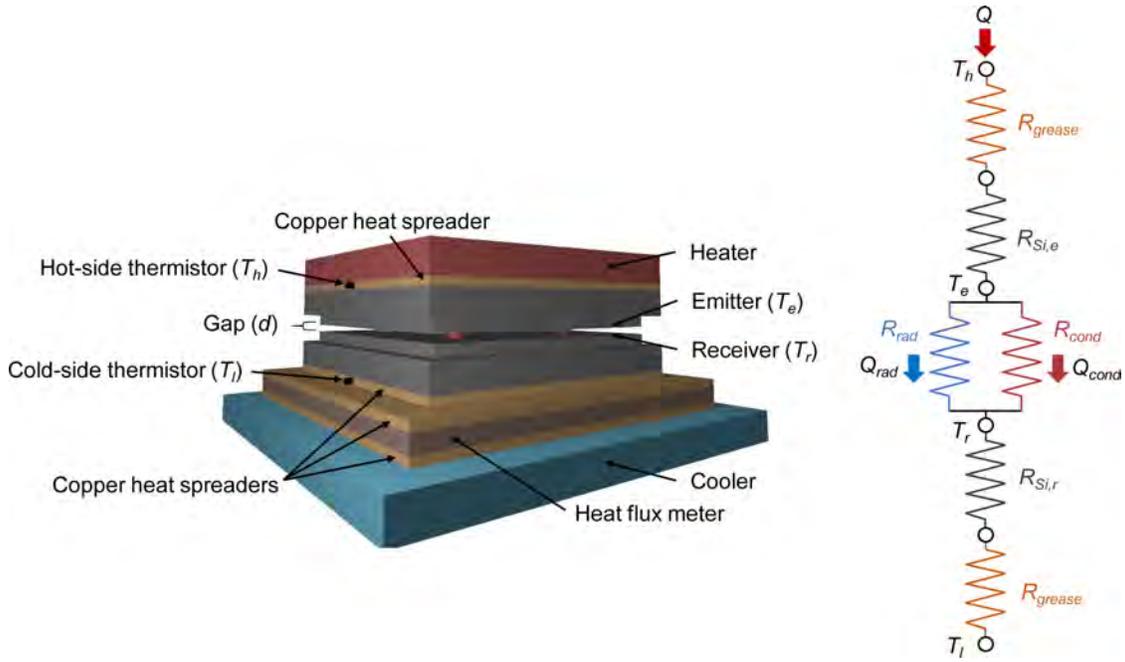

**Fig. 1 | Near-field radiative heat transfer (NFRHT) device and measurement setup. a**, The NFRHT device consists of an emitter and receiver, both made of doped silicon (Si), separated by a vacuum gap spacing, $d$, supported by SU-8 3005 micropillars. The micropillars are fabricated within ~ 4.5-μm-deep, ~ 215-μm-diameter pits etched into the emitter (see microscope image taken from Keyence VHX-5000). While the bottom view of the emitter shows four micropillars/pits (2 × 2 array), devices with 3 × 3 micropillar/pit arrays have also been tested. The micropillar diameters are 30 μm and 20 μm for the 2 × 2 and 3 × 3 arrays, respectively, resulting in equivalent total micropillar cross-sectional area for all devices. The vacuum gap separating the emitter and receiver surfaces corresponds to the difference between the height of the micropillars and the depth of the pits ($d = t_{mp} - t_{pit}$). Note that the NFRHT device schematic is not to scale. **b**, Imaging of a device gap spacing ($d \approx 380$ nm) via scanning electron microscopy (SEM, FEI Quanta 600 FEG). Each of the SEM image corresponds to a corner of the device. **c**, Heat transfer measurement setup and equivalent thermal circuit of the device. The setup from top to bottom consists of a thermoelectric heater, a hot-side thermistor embedded in a copper heat spreader for measuring the high temperature, $T_h$, the NFRHT device, a cold-side thermistor embedded in a copper heat spreader for measuring the low temperature, $T_l$, a heat flux meter surrounded by copper heat spreaders to ensure a uniform flux through the meter, and a thermoelectric cooler. The equivalent thermal circuit shows that the heat rate flowing through the device, $Q$, is the sum of heat rates due to conduction though the micropillars, $Q_{cond}$, and radiation between the emitter and receiver, $Q_{rad}$. The emitter and receiver temperatures adjacent to the vacuum gap, $T_e$ and $T_r$, are retrieved using the thermal resistances due to the thermal grease, $R_{grease}$, and the thermal resistances due to conduction within the doped Si emitter and receiver, $R_{Si,e}$ and $R_{Si,r}$.

**a**

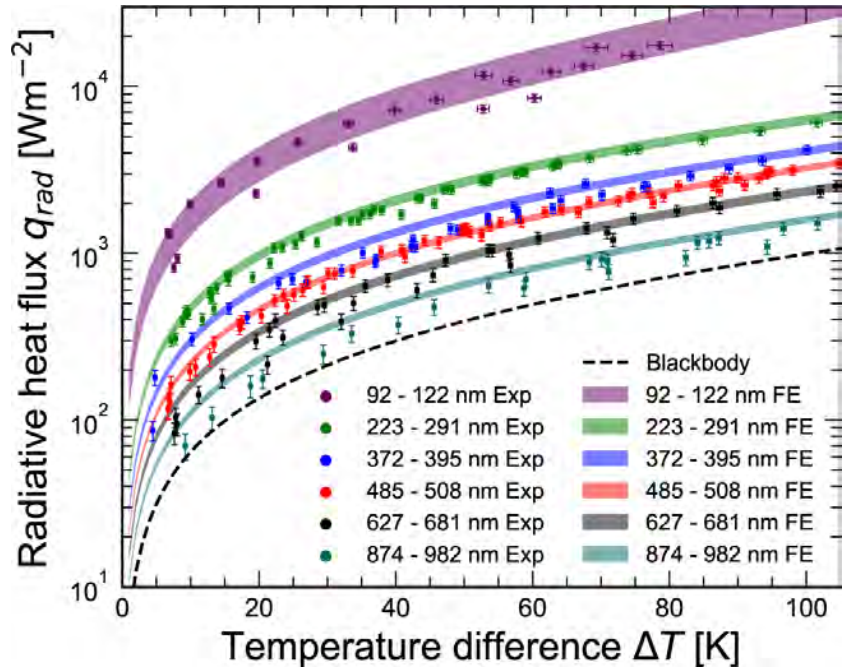

**b**

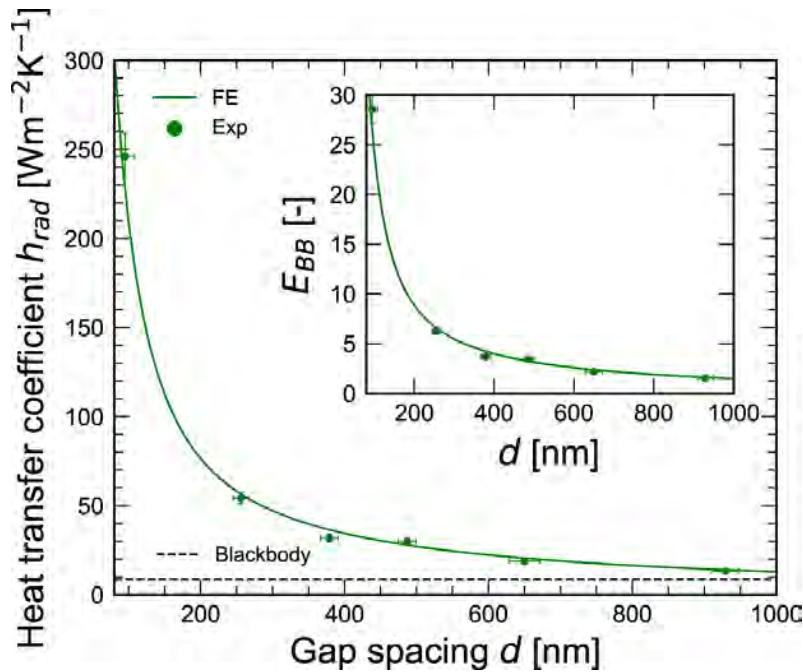

**Fig. 2 | Gap- and temperature-dependent radiative heat flux and heat transfer coefficient. a**, Radiative heat flux, $q_{rad}$, as a function of the temperature difference, $\Delta T$ ($= T_e - T_r$), where $T_r = 300 \pm 0.5$ K for six different devices with gap spacings, $d$, ranging from approximately 1000 nm down to 110 nm. The symbols display the experimental radiative flux, where conduction heat transfer through the micropillars and radiation heat transfer from the recessed areas (pits and

frame) are subtracted. The colored bands show theoretical predictions calculated via fluctuational electrodynamics (FE). The devices characterized by vacuum gap spacings of 372-395 nm and 485-508 nm have $3 \times 3$ micropillar/pit arrays while the other devices have $2 \times 2$ micropillar/pit arrays. **b**, Heat transfer coefficient due to radiative transfer between the emitter and receiver, $h_{rad}$, and enhancement with respect to the far-field blackbody limit, $E_{BB}$, as a function of the vacuum gap spacing, $d$, for a temperature difference, $\Delta T$, of $70 \pm 2$ K. The heat transfer coefficient for all devices exceeds the blackbody limit. A maximum enhancement of approximately 28.5 over the blackbody limit is measured for the device with the smallest vacuum gap spacing (92-122 nm). The gap spacing in panel **b** is determined by matching the overall device radiative flux, predicted via FE and the Derjaguin approximation, to a single, effective vacuum gap spacing.

**a**

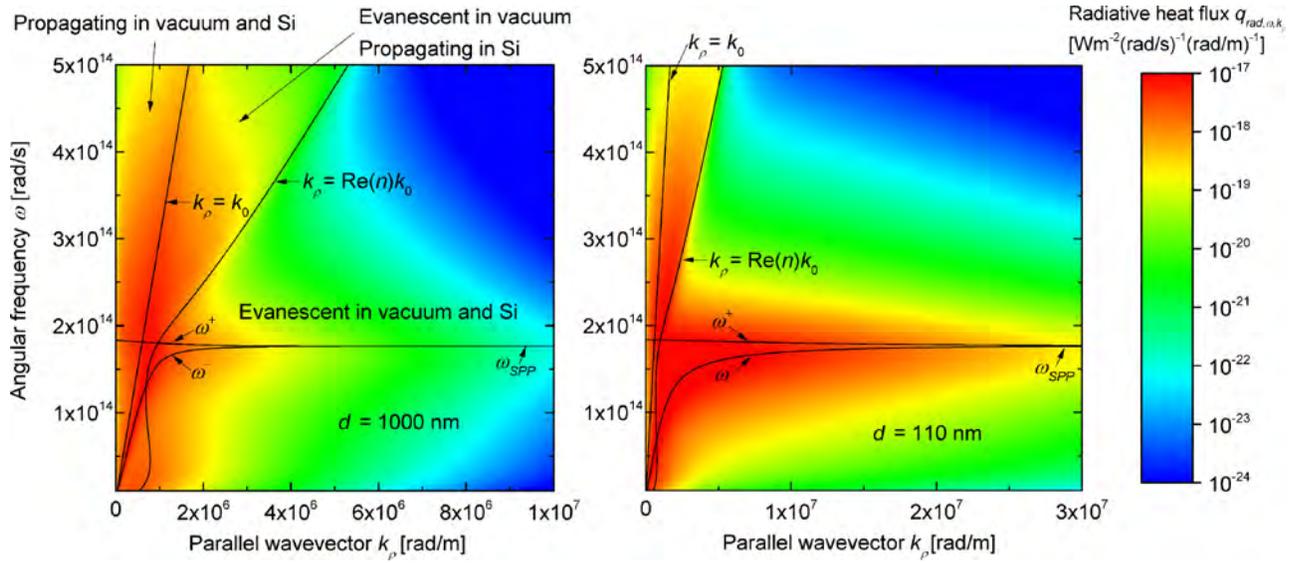

**b**

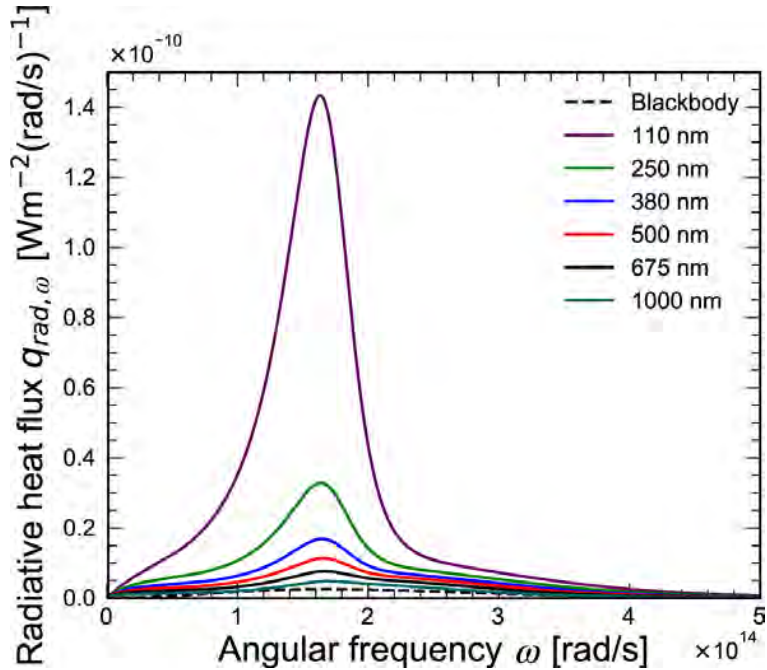

**Fig. 3 | Analysis of near-field radiative heat transfer enhancement. a,** Radiative heat flux, $q_{rad,\omega,k_\rho}$, per unit angular frequency, $\omega$, and parallel wavevector, $k_\rho$, for gap spacings of 1000 nm and 110 nm and a temperature difference, $\Delta T$, of 70 K. These gap spacing values are representative of the device characterized by the largest (874-982 nm) and smallest (92-122 nm) gap spacing. The region where $k_\rho$ is smaller than $k_0$ (= $\omega/c_0$) corresponds to modes propagating in both silicon (Si) and vacuum. The region defined by $k_0 < k_\rho < \mathrm{Re}(n)k_0$, where $n$ is the refractive

index of doped Si, describes frustrated modes that are propagating in Si but evanescent in vacuum. Surface plasmon-polaritons (SPPs), existing in the region where $k_\rho > \mathrm{Re}(n)k_0$, are evanescent in both Si and vacuum. The dispersion relation of SPPs in the Si-vacuum-Si configuration splits into antisymmetric, $\omega^+$, and symmetric, $\omega^-$, modes converging to the resonant frequency of a Si-vacuum interface, $\omega_{SPP}$. While the portion of the $\omega^+$ and $\omega^-$ branches existing in the region where $k_\rho < \mathrm{Re}(n)k_0$ satisfies the dispersion relation given by Eqs. (4) and (5), surface polaritons only exist in the region where $k_\rho > \mathrm{Re}(n)k_0$. **b,** Radiative heat flux, $q_{rad,\omega}$, per unit angular frequency, $\omega$, for all six devices. Calculations are performed at representative gap spacings of 110 nm, 250 nm, 380 nm, 500 nm, 675 nm, and 1000 nm for a temperature difference, $\Delta T$, of 70 K. For comparison, the radiative heat flux between two blackbodies is also plotted.

# Supplementary Information

# Near-field radiative heat transfer devices

John DeSutter[1], Lei Tang[2] and Mathieu Francoeur[1,*]

[1]Radiative Energy Transfer Lab, Department of Mechanical Engineering, University of Utah, Salt Lake City, UT 84112, USA

[2]Department of Mechanical Engineering, University of California, Berkeley, Berkeley, CA 94720, USA

<u>Supplementary Information Contents:</u>

**Supplementary Fig. 1 | Device fabrication.**

**Supplementary Fig. 2 | Heat flux meter (HFM) calibration.**

**Supplementary Fig. 3 | Gap spacing images obtained from scanning electron microscopy (SEM).**

**Supplementary Fig. 4 | Corner-to-corner room temperature bow of a device receiver.**

**Supplementary Fig. 5 | Two-dimensional (2D) room temperature bow of the smallest gap device receiver.**

**Supplementary Fig. 6 | Temperature-dependent bow of a doped silicon (Si) wafer.**

**Supplementary Fig. 7 | Gap- and temperature-dependent total heat rate.**

**Supplementary Fig. 8 | Temperature-dependent total heat rate with doped silicon (Si) surfaces in contact.**

**Supplementary Fig. 9 | Emitter deflection in the heat transfer measurement setup.**

**Supplementary Section 1: Calibration of heat flux meter and estimation of thermal grease resistance.**

**Supplementary Section 2: Gap spacing estimation.**

**Supplementary Section 3: Dielectric function of doped Si.**

---

[*]email: mfrancoeur@mech.utah.edu

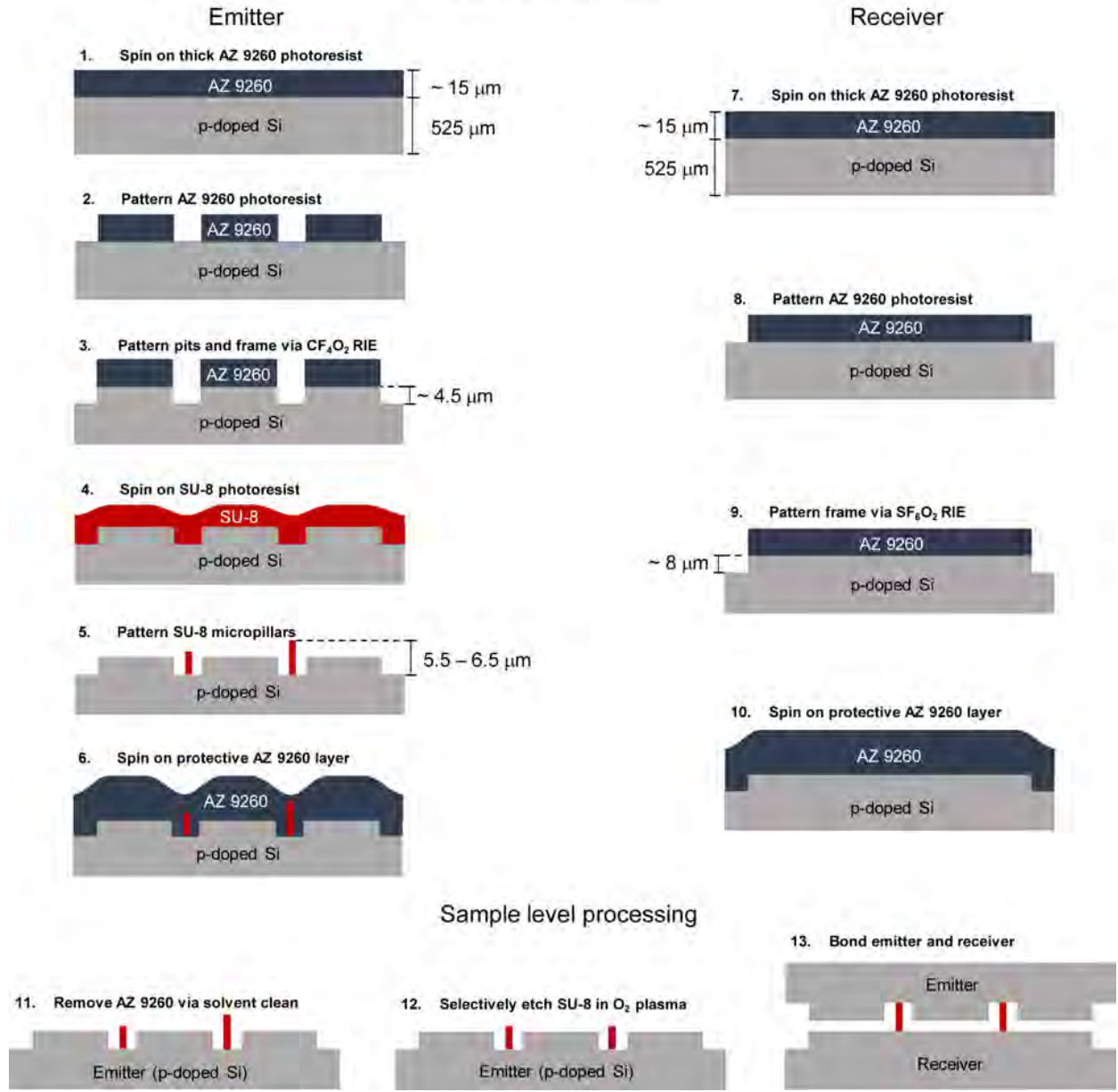

**Supplementary Fig. 1 | Device fabrication.** Main fabrication steps for wafer level processing (emitter and receiver) and sample level processing.

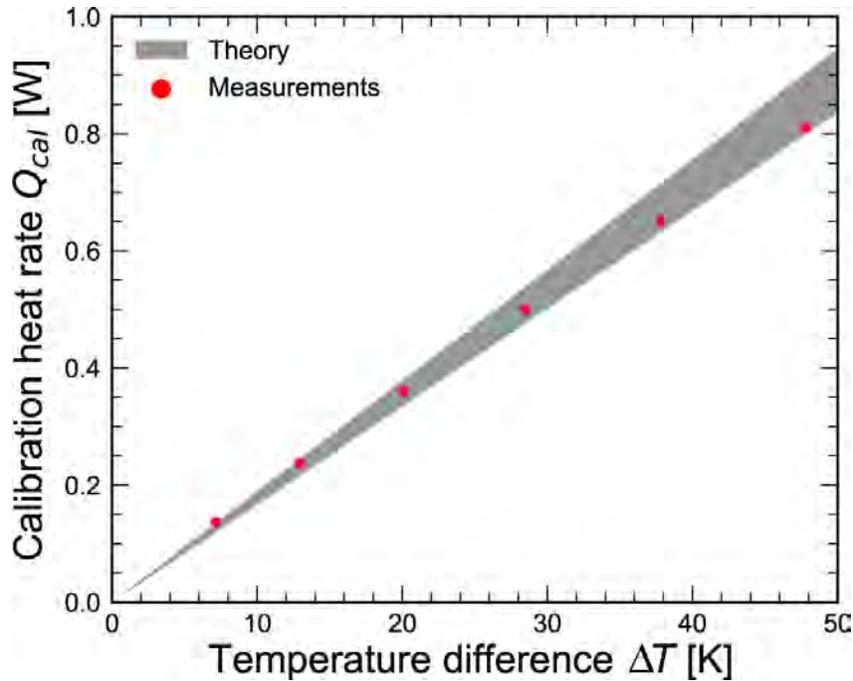

**Supplementary Fig. 2 | Heat flux meter (HFM) calibration.** Calibration heat rate by conduction, $Q_{cal}$, through a 1.1-mm-thick, $5 \times 5$ mm$^2$ soda-lime glass sample having a thermal conductivity of 0.94 Wm$^{-1}$K$^{-1}$ as a function of the temperature difference, $\Delta T = T_h - T_l$ ($T_l$ is maintained at $\sim$ 300 K). The symbols display experimental heat rate. The colored band shows theoretical predictions calculated by assuming one-dimensional, steady-state conduction heat transfer. The theoretical total thermal resistance includes the theoretical thermal resistance by conduction through the soda-lime glass sample and the experimentally determined thermal grease resistance. The band for theoretical predictions comes from the uncertainty associated with thermal grease resistance. Theoretical and experimental results are in good agreement, thus suggesting that the manufacturer supplied HFM sensitivity of 0.276 μV/(Wm$^{-2}$) is appropriate.

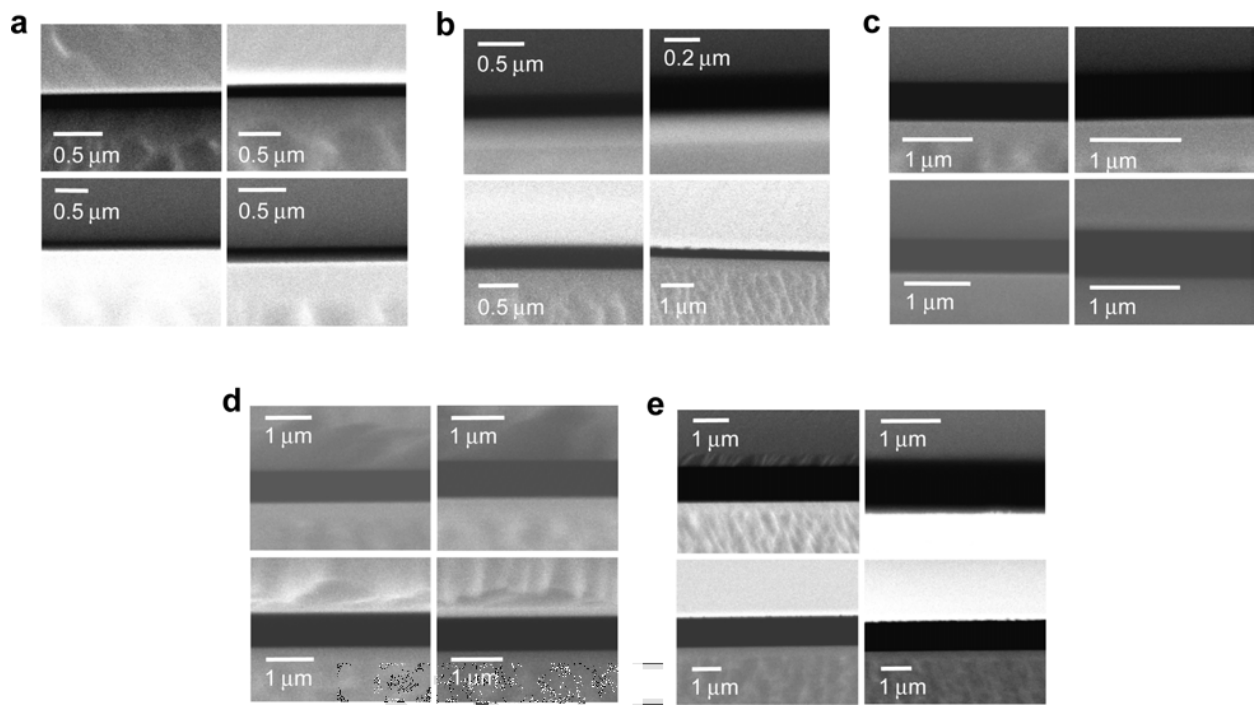

**Supplementary Fig. 3 | Gap spacing images obtained from scanning electron microscopy (SEM). a**, 92-122 nm device. **b**, 223-291 nm device. **c**, 485-508 nm device. **d**, 627-681 nm device. **e**, 874-982 nm device. In a specific panel, each of the SEM images corresponds to a different corner of a single device.

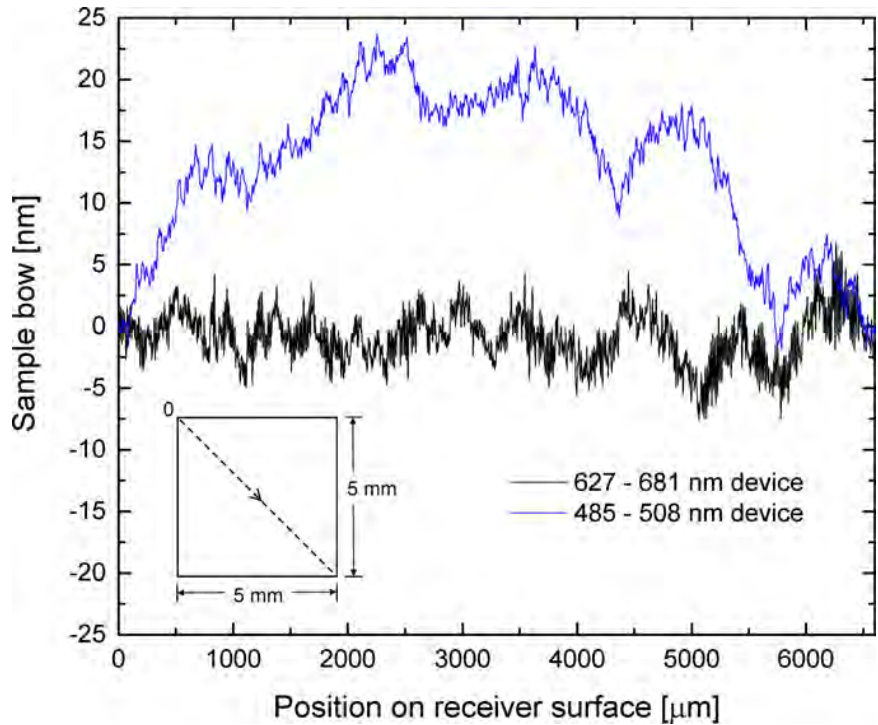

**Supplementary Fig. 4 | Corner-to-corner room temperature bow of a device receiver.** The bow of the doped silicon (Si) substrates used for fabricating near-field radiative heat transfer (NFRHT) devices is smaller than ~ 25 nm (measured using a Tencor P-20H profilometer). The receiver bows plotted correspond to the 627-681 nm (black) and 485-508 nm (blue) devices. The emitter and receiver bow for all other devices falls within this range.

**a**

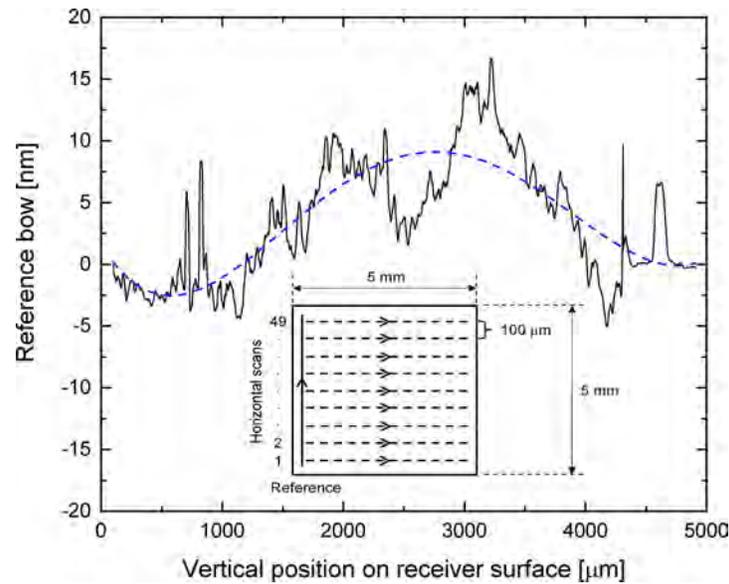

**b**

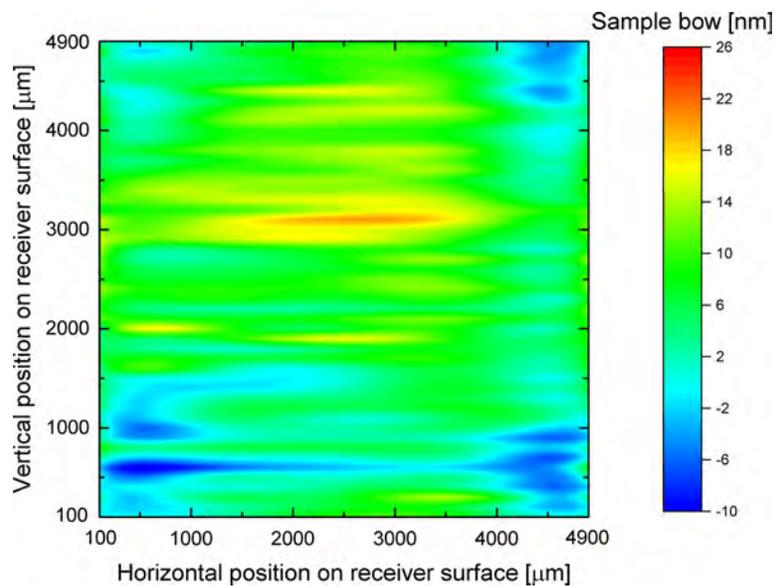

**Supplementary Fig. 5 | Two-dimensional (2D) room temperature bow of the smallest gap device receiver. a**, 2D topographic mapping of doped silicon (Si) substrates is obtained by performing 50 profilometry scans (measured using a Tencor P-20H profilometer). One vertical reference scan at a horizontal position of ~ 100 μm and 49 horizontal scans ranging from vertical position of ~ 100 μm to ~ 4900 μm constitute the 2D topographic mapping. **b**, 2D topographic mapping of the smallest gap device receiver. The bow of the doped Si substrates used for fabricating NFRHT devices is smaller than ~ 25 nm. The emitter and receiver bow for all other devices falls within this range.

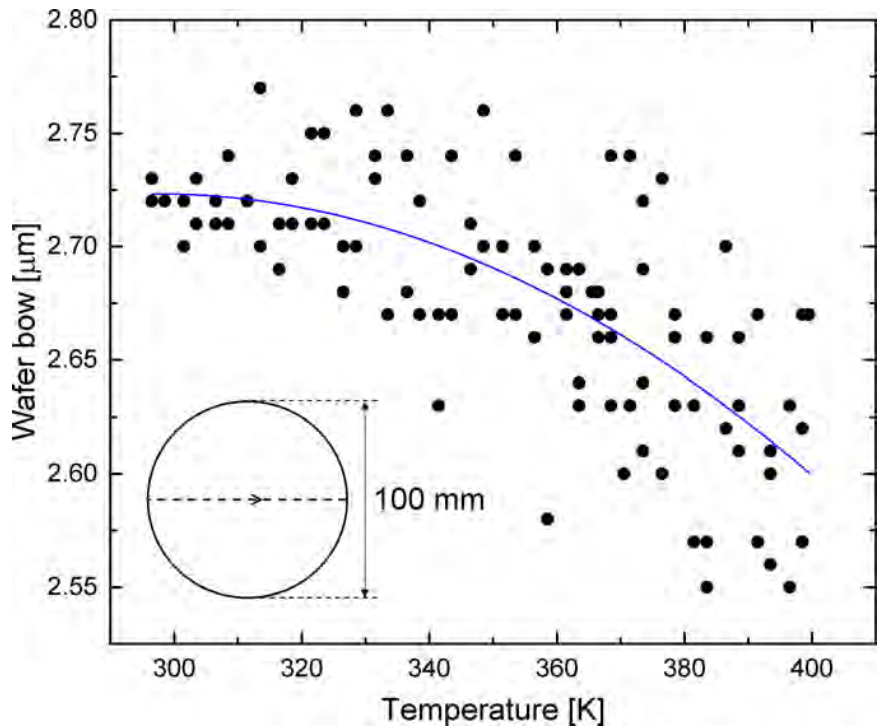

**Supplementary Fig. 6 | Temperature-dependent bow of a doped silicon (Si) wafer.** The room temperature bow of the 100-mm-diameter doped Si wafer used for fabricating near-field radiative heat transfer (NFRHT) devices is ~ 2.72 μm (measured using a Tencor FLX 2320), which is in good agreement with the data provided by the manufacturer (< 4.5 μm). The wafer bow slightly decreases to a value of ~ 2.6 μm as the temperature increases to ~ 400 K. This corresponds to a bow variation of ~ 4.5%. Variation of bow as a function of temperature is therefore negligible when calculating the radiative flux for the smallest gap device.

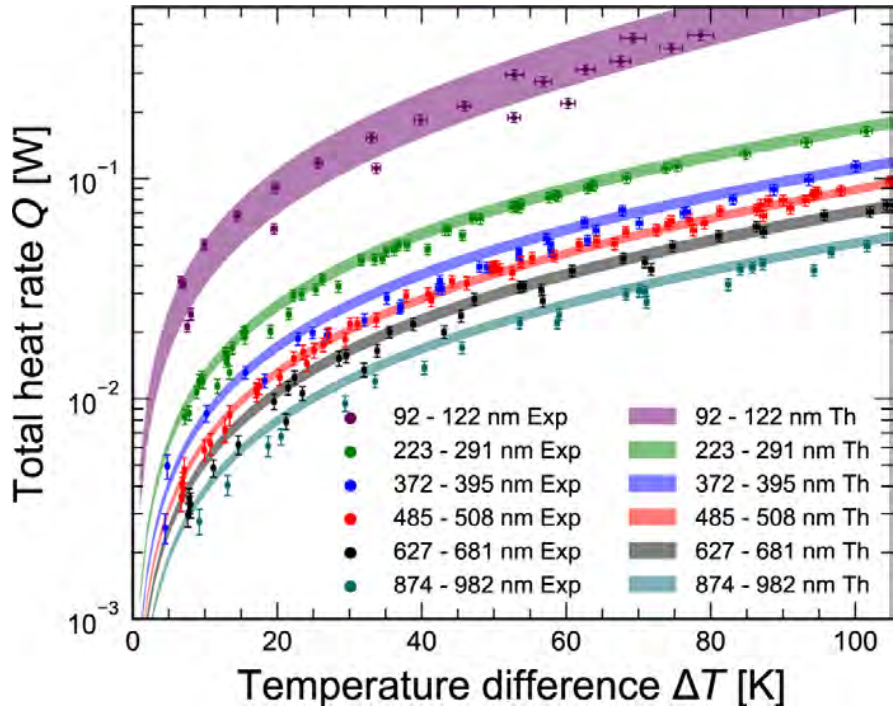

**Supplementary Fig. 7 | Gap- and temperature-dependent total heat rate.** Total heat rate, $Q$, as a function of the temperature difference, $\Delta T$ (= $T_e - T_r$), where $T_r$ = 300 ± 0.5 K for six different devices with gap spacings, $d$, ranging from approximately 1000 nm down to 110 nm. The symbols display the experimental heat rate, that includes all contributions, namely radiation heat transfer across the gap spacing, conduction heat transfer through the micropillars, and radiation heat transfer from the recessed areas (pits and frame). The colored bands show theoretical predictions calculated via fluctuational electrodynamics (FE) and a one-dimensional, steady-state conduction model.

**a**

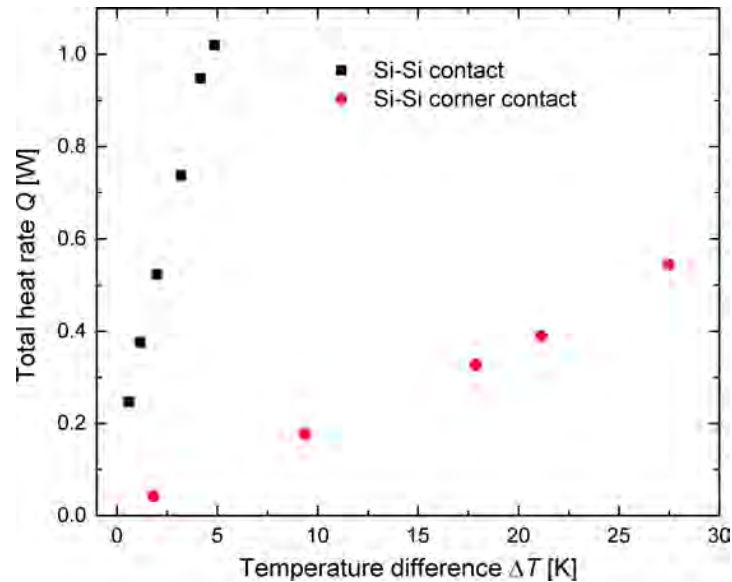

**b**

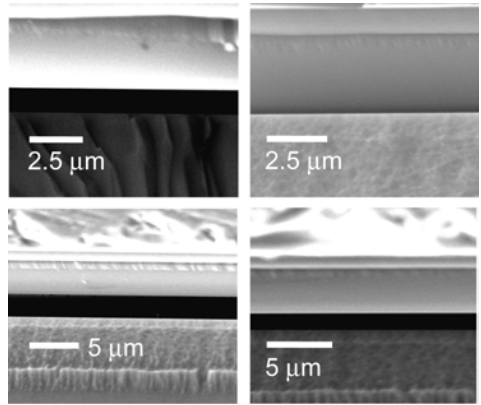

**Supplementary Fig. 8 | Temperature-dependent total heat rate with doped silicon (Si) surfaces in contact. a**, Total heat rate, $Q$, as a function of the temperature difference, $\Delta T$ (= $T_e - T_r$), for two Si surfaces in contact (i.e., no gap spacing) and two Si surfaces partially in contact. For a temperature difference of 5 K, the heat rate for two Si surfaces in contact exceeds the heat rate in the smallest gap device by a factor of ~ 50 (see Supplementary Fig. 7). **b**, Gap spacing images obtained from SEM of a device with one corner in contact (upper right image). Partial Si-Si contact is obtained by intentionally removing one micropillar. The three other gap spacings are 1.18 μm (upper left image), 2.31 μm (lower left image), and 1.26 μm (lower right image). For a temperature difference of 27.5 K, the heat rate between two Si surfaces partially in contact exceeds the heat rate in the largest and smallest gap devices by factors of ~ 55 and ~ 5, respectively. Therefore, the heat rate measured in the NFRHT devices cannot be due to Si-Si conduction. In all experiments, a 3 g mass is deposited on the heater to keep in place the different layers of the heat transfer measurement setup.

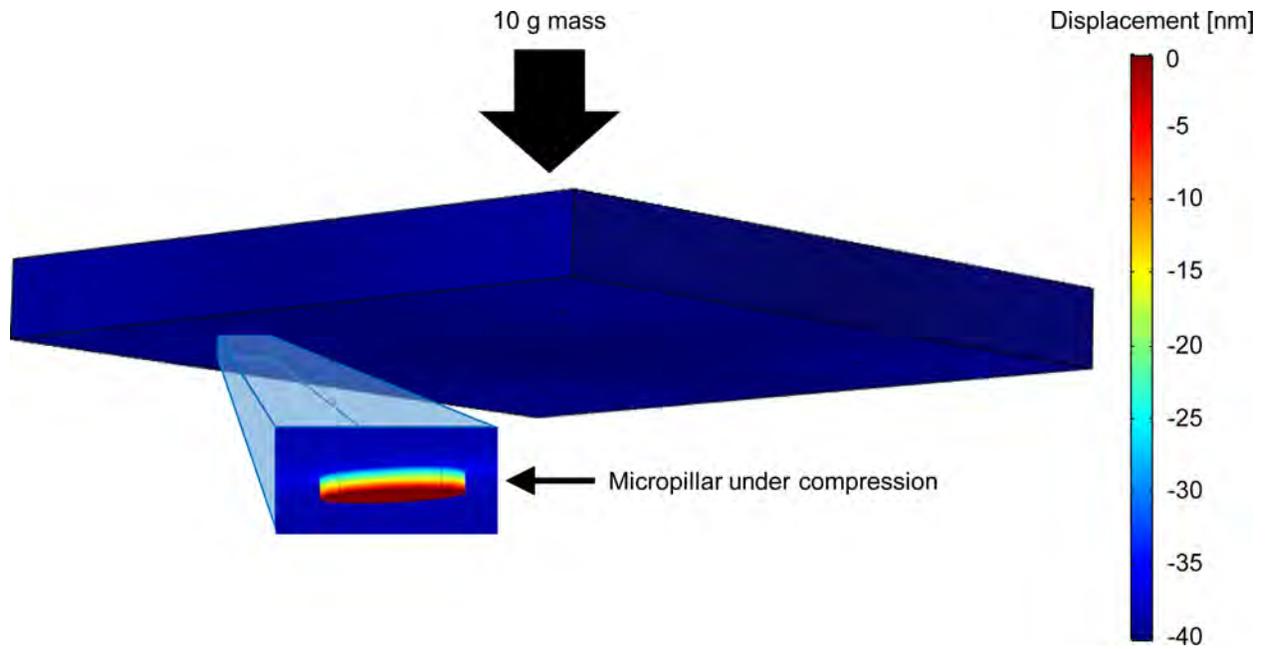

**Supplementary Fig. 9 | Emitter deflection in the heat transfer measurement setup.** A force due to a 10 g mass is applied on the top surface of the emitter. The lower faces of the four micropillars are held fixed (i.e., no displacement). The deflection of the emitter due to micropillar compression is uniform and takes a value ~ 40 nm when the emitter and receiver are both at 300 K. The magnified portion shows the displacement of a single micropillar due to compression.

**1. Calibration of heat flux meter and estimation of thermal grease resistance**

Heat transfer across the devices is measured with a $10 \times 10$ mm$^2$ heat flux meter (HFM), converting thermal energy into an electrical signal. An HFM is a thermopile where multiple thermocouples, separated by a material of low thermal conductivity, are connected in series. As heat flows through the HFM, a temperature difference across the thermocouples is induced thus generating a measureable voltage drop. The measured voltage drop is converted into a heat flux (Wm$^{-2}$) via the HFM sensitivity that has units of $\mu$V/(Wm$^{-2}$). The heat rate (W) is readily obtained by multiplying the heat flux by the surface area of the HFM. The sensitivity of the HFM provided by the manufacturer is 0.276 $\mu$V/(Wm$^{-2}$). A procedure similar to the one proposed by Watjen et al.[1] is used to calibrate the HFM.

HFM calibration requires knowledge of the thermal resistance due to thermal grease applied at the interfaces between the device and setup shown in Fig. 1c. Thermal grease resistance is experimentally determined by replacing the near-field radiative heat transfer (NFRHT) device by a 525-$\mu$m-thick, $5 \times 5$ mm$^2$ sample of silicon (Si) with boron doping of $\sim 4.6 \times 10^{19}$ cm$^{-3}$. Assuming one-dimensional, steady-state conduction heat transfer, the calibration heat rate can be written as:

$$Q_{cal} = \frac{T_h - T_l}{R_{tot}} \tag{S1}$$

where $R_{tot}$ is the total thermal resistance. This total thermal resistance includes the thermal grease resistance, $R_{grease}$, at the two copper-silicon interfaces, and the thermal resistance by conduction through Si, $R_{Si}$ ($R_{tot} = R_{Si} + 2R_{grease}$). The thermal conductivity of highly doped Si exceeds 100 Wm$^{-1}$K$^{-1}$,[2] thus making $R_{Si}$ negligible compared to that of the thermal grease (i.e., $R_{tot} \approx 2R_{grease}$). Based on 24 measurements with temperature differences from 3.4 to 15.5 K, $R_{grease}$ values ranging from 2.8 to 6.2 K/W are determined. To ensure good contact at the interfaces where thermal grease is applied, 10 g and 3 g masses are placed on the heater during the analysis. It is determined that thermal grease resistance is not noticeably impacted by the difference in mass.

HFM calibration is done by performing heat transfer measurements using a material having a known thermal conductivity. Specifically, a 1.1-mm-thick, $5 \times 5$ mm$^2$ soda-lime glass sample with a thermal conductivity of 0.94 Wm$^{-1}$K$^{-1}$, as specified by the manufacturer (Valley Design Corp),[3] is used. Doped Si is replaced in the setup shown in Fig. 1c by the soda-lime glass sample. The heat rate by conduction during calibration, $Q_{cal}$, is measured as a function of the temperature difference, $\Delta T = T_h - T_l$ ($T_l$ is maintained at 300 K), using the HFM sensitivity provided by the manufacturer. The experimental measurements are shown in Supplementary Fig. 2. The correctness of the HFM sensitivity is assessed by calculating the heat rate by conduction using Eq. (S1), where $R_{tot} = R_{glass} + 2R_{grease}$. Here, the theoretical thermal resistance through the soda-lime glass sample, $R_{glass}$, is not negligible with respect to the thermal grease resistance and takes a value of 46.8 K/W. Using the thermal grease resistance, $R_{grease}$, experimentally estimated with the doped Si sample, the total thermal resistance ranges from 52.4 to 59.2 K/W. Theoretical predictions of $Q_{cal}$ are also plotted in Supplementary Fig. 2 as a function of the temperature difference. Clearly, experimental data fall within the theoretical heat rate range. It is therefore concluded that the sensitivity of 0.276 µV/(Wm$^{-2}$) provided by the manufacturer is correct. As such, this sensitivity value is used in all NFRHT experiments.

It is worth noting that thermal grease resistance is small compared to the radiative thermal resistance across the vacuum gap spacing in the NFRHT devices. For a temperature difference of 70 K, the radiative thermal resistances for the smallest and largest gap devices are ~ 180 K/W and ~ 2860 K/W, respectively.

## 2. Gap spacing estimation

The NFRHT device structural integrity enables gap spacing visualization via scanning electron microscopy (SEM). This is achieved by adhering the device to a vertical mount inside the SEM chamber such that the gap spacing is clearly exposed to the electron gun. Two corners are imaged before the device is removed and rotated by 180°. The device is then placed again in the chamber to image the other two corners. Visualizing all four corners of a device is crucial, as potential particle contamination prior to bonding can cause gap spacing variation exceeding 1 μm. Gap spacing SEM images of the six devices analyzed in this work are provided in Fig. 1b and Supplementary Fig. 3. The ability to mount a device vertically and remove it from adhesive tape emphasizes the robustness of the NFRHT devices. It is worth mentioning that when a device is removed from the heat transfer measurement setup (see Fig. 1c), the adhesion of the thermal grease causes the emitter to pull apart from the receiver. To ensure that the devices are not failing during heat transfer measurements due to thermal expansion of the emitter, two test devices with gap spacings ~ 1200 nm and the same micropillar area as the six measured devices have been placed in the setup without using thermal grease on the emitter side. Temperature differences exceeding 115 K where applied to both devices while the receiver was held at ~ 300 K. In both cases, the devices remained intact.

A force onto the heater is applied via calibrated masses in order to minimize thermal contact resistances and to keep in place the different layers in the heat transfer measurement setup. A 10 g mass was used for the 92-122 nm, 485-508 nm, 627-681 nm and 874-982 nm devices, while a 3 g mass was used for the 223-291 nm and 372-395 nm devices. The force exerted on the NFRHT device may, however, impact the SU-8 micropillar height, thus potentially affecting the gap spacing, $d$, due to deflection of the emitter with respect to the receiver. Young's modulus of SU-8 is in the range of ~ 3.5 to 4.1 GPa at room temperature[4]. Using a Young's modulus of 3.8 GPa, a one-dimensional linear elastic analysis suggests that the micropillars compress by ~ 43 nm when a 10 g mass is applied. When both emitter and receiver are at 300 K, COMSOL simulations reveal that the SU-8 micropillars compress by ~ 40 nm (see Supplementary Fig. 9), which is in excellent agreement with the analytical result. When the emitter is at a temperature of 380 K, it is estimated that micropillar compression increases to ~ 50 nm using temperature-dependent mechanical properties of SU-8[5]. Note that this temperature-dependent analysis also considers thermal expansion of SU-8 and doped Si. A similar analysis has been performed for the devices subject to the 3 g mass, although the effect is smaller. Here, the micropillar compression is only ~ 12 nm when both the emitter and receiver are at room temperature, and decreases to 0.3 nm when the emitter temperature is increased at 400 K due to thermal expansion. In addition, COMSOL simulations of the bonded devices suggest that the emitter deflection with respect to the receiver is uniform across the entire surface regardless of the emitter temperature.

In Fig. 2a, the gap spacing range specified for a given device (e.g., 92-122 nm for the smallest gap device) comes from SEM images of the four corners of the device (132-162 nm) and the deflection of the emitter with respect to the receiver due to SU-8 micropillar compression at 300 K (40 nm). Gap spacing variation due to emitter and receiver bow (less than ~ 25 nm; see Supplementary Figs. 4 and 5) is considered only when calculating the radiative flux for the smallest gap device, as discussed in Methods. It is also important to note that the temperature-dependence of the SU-8 micropillar compression is taken into account when calculating radiation and conduction heat transfer between the emitter and receiver. Therefore, the theoretical results reported in Fig. 2 and Supplementary Fig. 7 fully account for the gap spacing variation as a function of the emitter temperature due to temperature-dependent SU-8 micropillar behavior.

## 3. Dielectric function of doped Si

The doping- and temperature-dependent dielectric function for boron-doped Si is calculated using the method proposed by Basu et al.[6,7]. Since highly doped Si has metallic behavior, its dielectric function is described by a Drude model:

$$\varepsilon(\omega) = \varepsilon_\infty - \frac{\omega_p^2}{\omega(\omega + i\gamma)} \tag{S2}$$

where $\varepsilon_\infty = 11.7$ is the limiting value of the dielectric function at high frequency, $\omega_p$ is the plasma frequency and $\gamma$ is the scattering rate. Both $\omega_p$ and $\gamma$ are dependent on the boron doping concentration and temperature, and are given by:

$$\omega_p = \sqrt{\frac{N_h e^2}{m^* \varepsilon_0}} \tag{S3}$$

$$\gamma = \frac{e}{m^* \mu} \tag{S4}$$

where $N_h$ is the hole concentration, $e$ is the electron charge, $\varepsilon_0$ is the permittivity of vacuum, and $m^* = 0.37 m_0$ is the hole effective mass where $m_0$ is the free electron rest mass in vacuum. The temperature-dependent mobility, $\mu$, is defined as:

$$\mu(T) = \Phi^{1.5} \left( \mu_1 \exp(-p_c / N_h) + \frac{\mu_{\max}}{1 + (N_h / C_r)^\alpha} - \frac{\mu_2}{1 + (C_s / N_h)^\beta} \right) \tag{S5}$$

where $\mu_1 = 44.9$ cm$^2$/Vs, $\mu_{max} = 470.5$ cm$^2$/Vs, $\mu_2 = 29.0$ cm$^2$/Vs, $C_r = 2.23 \times 10^{17}$ cm$^{-3}$, $C_s = 6.10 \times 10^{20}$ cm$^{-3}$, $\alpha = 0.719$, $\beta = 2$, $p_c = 9.23 \times 10^{16}$ cm$^{-3}$, and $\Phi = T/300$ is a reduced temperature in kelvin. The hole concentration, $N_h$, is determined from the boron doping concentration, $N_A$, as follows:

$$\frac{N_h}{N_A} = 1 - A \exp\{-[B \ln(N_A / N_0)]^2\} \tag{S6}$$

where $A = 0.2364\Phi^{-1.474}$, $N_0 = 1.577 \times 10^{18}\Phi^{0.46}$, and $B = 0.433\Phi^{0.2213}$ if $N_A < N_0$ and $B = 1.268 - 0.338\Phi$ if $N_A > N_0$.